\documentclass[amstex,aps,showpacs,preprint]{revtex4}%
\usepackage{amsmath}
\usepackage{graphicx}%
\usepackage{amsfonts}%
\usepackage{amssymb}

\begin{document}

\title{Can Bohmian trajectories account for quantum recurrences having classical periodicities?}
\author{A.\ Matzkin}
\affiliation{Laboratoire de Spectrom\'{e}trie physique (CNRS
Unit\'{e} 5588), Universit\'{e} Joseph-Fourier Grenoble-1, BP 87,
38402 Saint-Martin, France}

\begin{abstract}
Quantum systems in specific regimes display recurrences at the
 period of the periodic orbits of the corresponding
classical system. We investigate the excited hydrogen atom in a
magnetic field -- a prototypical system of 'quantum chaos' -- from
the point of view of the de Broglie Bohm (BB) interpretation of
quantum mechanics. The trajectories predicted by BB theory are
computed and contrasted with the time evolution of the wavefunction,
which shows pronounced features at times matching the period of
closed  orbits of the \emph{classical} hydrogen in a magnetic field
problem. \emph{Individual} BB trajectories do not possess these
periodicities and cannot account for the quantum recurrences. These
recurrences can however be explained by BB theory by considering the
\emph{ensemble} of trajectories compatible with an initial
statistical distribution, although none of the trajectories of the
ensemble are periodic, rendering unclear the dynamical origin of the
classical periodicities.
\end{abstract}


 \maketitle

\section{Introduction}

The manifestation of classical orbits has been found in a host of
quantum systems, displaying features such as scars of wavefunctions
along periodic orbits of the corresponding classical system or time
recurrences appearing at the periods of classical closed orbits
\cite{brack badhuri}. These features have been observed
experimentally in systems such as mesoscopic devices or atoms in
external fields. From within a pure Schr\"{o}dinger based approach,
these phenomena may appear as coming out of the blue. They are
however well understood by performing asymptotic  $\hbar$
expansions. In particular it is straightforward to show (eg,
\cite{grosche}) that the evolution operator obtained from the path
integral expression becomes to first order in $\hbar$

\begin{equation}
\left\langle x_{2},t_{2}\right|  e^{-iH(t_{2}-t_{1})/\hbar}\left|  x_{1}%
,t_{1}\right\rangle =\left(  2i\pi\hbar\right)  ^{-D/2}\sum_{k}\left|
\det\frac{\partial^{2}R_{k}}{\partial x_{2}\partial x_{1}}\right|  ^{1/2}%
\exp\frac{i}{\hbar}\left[  R_{k}(x_{2},x_{1};t_{2}-t_{1})-\phi_{k}\right]
+O(\hbar). \label{1}%
\end{equation}
We have assumed for simplicity a time-independent Hamiltonian $H$ in
$D$ dimensional configuration space. The sum runs on the
\emph{classical paths} $k$ connecting $x_{1}$ and $x_{2}$ and
$R_{k}$ is the classical action along the trajectory $k;$ it
satisfies the Hamilton-Jacobi equation of classical
mechanics \cite{goldstein}%
\begin{equation}
\frac{\partial R(x,t)}{\partial t}+\frac{(\triangledown R(x,t))^{2}}%
{2m}+V(x)=0. \label{2}%
\end{equation}
The determinant is linked to the classical density and $\phi_{k}$ is
an additional phase that keeps track of the points where the
classical amplitude is singular.\ The physical meaning of Eq.
(\ref{1}) is simple: when $\hbar/R$ is small (a situation to be
termed here 'semiclassical regime') propagation in configuration
space takes place only along the classical paths, the sum reminding
us that the \emph{wave} takes all the \emph{paths} simultaneously
with a given weight -- the classical amplitude.

An alternative interpretation of quantum phenomena hinges on the
existence of point-like particles following a well-defined
space-time trajectory -- a \emph{quantum trajectory}. The de Broglie
Bohm (BB) theory is by far the best-known and most developed of
hidden-variables theories, and BB\ trajectories have been computed
for a wide range of quantum systems (see \cite{holland93} and Refs
therein as well as more recent work e.g.
\cite{alcantara98,nogami00,wisniacki03,matz05}. One of the main
motivations behind the BB theory is to bridge the gap between
classical and quantum mechanics.\ Indeed the interpretation of
quantum phenomena by way of a statistical distribution of particles
moving along well-defined quantum trajectories appears as an
attractive manner of understanding how classical mechanics can
emerge from quantum phenomena.

The main concern of this work is to analyze the role of quantum
trajectories as predicted by the de Broglie-Bohm interpretation in
quantum systems displaying the fingerprints of classical
trajectories. In such systems, the wavefunction is carried by
classical trajectories, and it is therefore of interest to compare
and contrast classical and quantum trajectories. This will be done
for a well known prototypical system, an excited hydrogen atom in a
magnetic field \cite{friedrich wintgen}. This system has been
heavily investigated, both theoretically and experimentally, in the
past 20 years and the success of its semiclassical analysis has
converted this sytem into a paradigm of ''quantum chaos'' . We will
briefly present the main characteristics of this system in Sec.\ 2.
We will then summarize the main properties of BB trajectories and
their expected behaviour in the semiclassical regime.\ Specific
quantum trajectories for the hydrogen atom in a magnetic field will
be computed in Sec\ 4. We will see that observable quantum
recurrences are ruled by the periodicity of the periodic orbits of
the corresponding classical system; the role of the quantum
trajectories in accounting for the recurrences will be discussed in
Sec\ 5.

\section{The hydrogen atom in a magnetic field}

The Hamiltonian describing the hydrogen atom in a magnetic field is given by
(eg the review paper \cite{friedrich wintgen})%
\begin{equation}
H=\frac{p^{2}}{2m} + \left( \frac{
qBL_{z}}{2mc}-\frac{q^{2}}{r}+\frac{(x^{2}+y^{2})q^{2}B^{2}}{8mc^{2}}\right)
, \label{5}%
\end{equation}
where $B$ is the strength of the magnetic field oriented in the $z$
direction, $\ m$ the mass of the electron, and $r$ the distance of
the electron relative to the nucleus. The spherical symmetry of the
Coulomb field is broken by the magnetic field, leaving an axial
symmetry (invariance around the $z$ axis).\ We will take $L_{z}=0$
in what follows and we will assume $B$ is sufficiently strong so
that perturbation theory is not necessarily valid. It can be shown
that $H$ possesses a scaling property, from which it follows that
the classical dynamics does not depend independently on the values
of the energy $E$ and of the intensity $B$ of the field but on the
ratio $\epsilon =EB^{-2/3}$ known as the scaled energy.\ For
$\epsilon\rightarrow-\infty$ the dynamics is near-integrable whereas
phase space is fully chaotic for $\epsilon\gtrsim-0.1$ and of mixed
nature for $-0.8\gtrsim\epsilon\gtrsim-0.1$.

\begin{figure}[tb]
\begin{center}
\includegraphics[height=2.1in,width=3in]{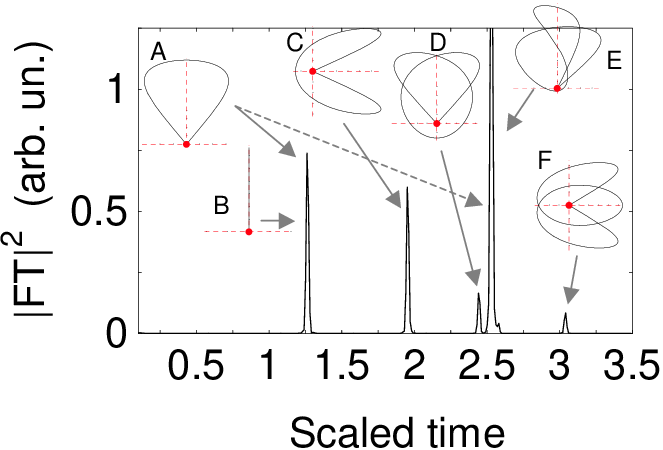}
\caption[]{Recurrence spectrum of the hydrogen atom in a magnetic
field obtained from quantum computations (Fourier transform of the
computed photoabsoption spectrum excited from the ground state with
the laser polarized along the field axis). The excitation energy
spans the interval $77<n<155$ and the magnetic field varies
accordingly so that the scaled energy stays fixed at
$\epsilon=-0.3$, thereby obtaining a \emph{scaled} spectrum. The
diagrams show the shape of the orbits closed at the nucleus of the
corresponding (scaled) classical problem in the $(\varrho,z)$ plane
($B$ is along the vertical $z$ axis). The arrows indicate the orbit
whose period matches the time of a given peak in the recurrence
spectrum. Note that more than one orbit can contribute to a given
peak, and that the repetitions of an orbit contribute to peaks
appearing at multiple integer times of the fundamental period (eg,
dotted arrow for the orbit A). } \label{fig1}
\end{center}
\end{figure}

The Schr\"{o}dinger equation, obtained from the standard
quantization of $H$, is simplified by eliminating the trivial
azimuthal angle.\ We are left with a nonseparable 2 dimensional
($\varrho,z$) problem which does not admit analytical solutions;
$\varrho,z$ are the rectangular (cylindrical) coordinates in the
axial plane. Obtaining the bound energies $E_n$ and the
eigenfunctions $\psi_{E_n}(\varrho,z)$ for highly excited states
therefore involves numerical computations with large basis sizes. We
will employ atomic units, the energies of the electron being labeled
by $E_n=-1/2n^2$ (where $n$ is of course not an integer). For small
$\epsilon$ the energy eigenvalues follow the well known pattern
given by perturbation theory (Zeeman effect) but as $\epsilon$
increases the spectrum becomes very complex, as the spherical $n/L$
degeneracy of the free-field atom is totally broken and thousands of
energy levels appear.\ The interpretation of individual levels
becomes meaningless, but it was gradually realized that
well-resolved peaks are visible by taking a Fourier transform of the
photoabsorption spectrum (obtaining what is called a recurrence
spectrum).\ These peaks, related to the large scale fluctuations of
the spectrum, appear at times corresponding to periods of classical
orbits closed at the nucleus.

A typical computed recurrence spectrum involving photoabsorption
from the ground state of the hydrogen atom is given in Fig.\ 1.\
Sharp peaks are visible. Above each peak, we have drawn the shape of
the classical orbit whose period matches the recurrence time of the
peak. This plot arises from quantum calculations, but recurrence
spectra have been experimentally observed in hydrogen \cite{mainprl}
as well as other species of one electron ('Rydberg') atoms
\cite{delande94} and molecules \cite{matz gauyacq} in external
fields \footnote{The recurrence spectrum shown in Fig.\ 1\ arises
from the Fourier transform of a \emph{scaled-energy} photoabsorption
spectrum where both the energy $E$ and field $B$ are varied so as to
keep the scaled energy $\epsilon$ constant (in a standard spectrum,
$B$ is fixed and only $E$\ varies). This results in considerably
narrow peaks, instead of wide overlapping structures that would be
harder to resolve (most experiments reported in
\cite{mainprl,delande94} were performed employing scaled energy
spectroscopy techniques).}. Purely semiclassical calculations have
also been undertaken, reaching an excellent agreement with
experimental observations and exact quantum calculations.\ The
semiclassical formalism, known as 'Closed orbit theory'
\cite{delos88}, starts from the semiclassical propagator (\ref{1})
and explains the recurrences observed with classical periodicity by
the propagation of the laser excited electron waves along the
classical trajectories that start and end at the nucleus: every such
orbit produces a peak whose height depends on the classical
amplitude of the orbit.\ If several orbits have the same or nearly
the same period (as happens in Fig 1) the height of the peak depends
on the interference between the orbits, and the phases $\phi_{k}$ of
Eq. (\ref{1}) play a crucial role.

\section{De Broglie-Bohm trajectories}

The de Broglie-Bohm interpretation of quantum mechanics has become
increasingly popular in the last decade and excellent accounts of the theory
are available \cite{holland93,bohm hiley}. The main dynamical equations arise
from the polar decomposition of the wavefunction in configuration space. Put
\begin{equation}
\psi(\mathbf{r},t)=\rho(\mathbf{r},t)\exp(i\sigma(\mathbf{r},t)/\hbar)
\label{e30}%
\end{equation}
where $\rho$ and $\sigma$ are real functions. The Schr\"{o}dinger equation
becomes equivalent to the coupled equations%
\begin{align}
\frac{\partial\sigma}{\partial t}+\frac{(\triangledown\sigma)^{2}}%
{2m}+V-\frac{\hbar^{2}}{2m}\frac{\triangledown^{2}\rho}{2\rho}  &
=0\label{e33}\\
\frac{\partial\rho^{2}}{\partial t}+\triangledown(\rho^{2}\triangledown
\sigma/m)  &  =0. \label{e34}%
\end{align}
$\rho^{2}$ gives the statistical distribution of the particle (here the
electron) whereas the trajectory is obtained by integrating the equation of
motion%
\begin{equation}
\frac{d\mathbf{r}}{dt}=\frac{\mathbf{\triangledown}\sigma(\mathbf{r},t)}{m}
\label{e10}%
\end{equation}
where the initial position of the electron $\mathbf{r}(t=0)$ lies
within the initial distribution $\rho(t=0).$ $V$ is the external
potential due to the Coulomb and magnetic fields (term between
$(...)$ in Eq. (\ref{5})) whereas the last term in Eq. (\ref{e33})
acts as a state-dependent 'quantum' potential.

Despite the formal similarity between Eq. (\ref{e33}) and the
classical Hamilton-Jacobi Eq. (\ref{2}), it is well established that
generic quantum trajectories are highly nonclassical
\cite{holland93,appleby}.\ This is still true in the semiclassical
regime: the reason is that the nonrelativistic BB theory is grounded
on a hydrodynamic framework whereby the local probability density
current%
\begin{equation}
\mathbf{j}(\mathbf{r},t)=\frac{\hbar}{2mi}\psi^{\ast}(\mathbf{r}%
,t)\overleftrightarrow{\nabla}\psi(\mathbf{r},t)\label{15}%
\end{equation}
is linked to the velocity of the particle (\ref{e10}) by%
\begin{equation}
\mathbf{j}(\mathbf{r},t)=\rho^{2}(\mathbf{r},t)\mathbf{v}(\mathbf{r}%
,t).\label{16}%
\end{equation}
Quantum trajectories are thus tangent to the local flow, as required
since two space-time points must be linked by a single trajectory.\
In the semiclassical approximation however the classical
trajectories do not in general follow the flow -- rather the flow
results from the interfering average of different bits of the
wavefunction, each carried by a classical trajectory. Indeed if we
assume a wavefunction initially ($t=0$) localized at
$\mathbf{r}_{0}$, at later times we have in the semiclassical approximation%
\begin{equation}
\psi(\mathbf{r},t)=\sum_{k}A_{k}(\mathbf{r},\mathbf{r}_{0},t)\exp
i(R_{k}(\mathbf{r},\mathbf{r}_{0},t)-\phi_{k}),\label{18}%
\end{equation}
where $A_{k}$ includes the prefactor given in Eq. (\ref{1}) and
quantities relative to the initial wavefunction.\ By plugging in Eq.
(\ref{18}) into the expression for the probability density amplitude
(\ref{15}) and comparing with Eq. (\ref{16}), it is immediate to see
that if only a single classical trajectory contributes to the sum
(\ref{18}), we have $\mathbf{j}=\left| A\right|  ^{2}\triangledown
R/m$ and the quantum trajectory approaches the classical one.\ This
is a very restrictive condition: even in conservative one
dimensional systems a wavepacket will be carried by several
classical trajectories, each having a slightly different energy. The
resulting interferences will prevent the BB trajectories to follow
the classical ones, as examined in a previous work for radial
Rydberg wavepackets \cite{matz05} In the general multidimensional
case several (quite often an infinity of) trajectories can be
launched classically at a given energy from a given
$\mathbf{r}_{0}$, and an initial wavepacket will contain a high
number of energy eigenstates (depending on the form of the initial
wavefunction). $\mathbf{j}$ then contains a double sum involving
correlations (in the form of interference terms) between all these
classical orbits (see Sec\ 6.4\ of \cite{holland93} where an
analogue simple example is worked out). Hence, quantum trajectories
are not expected to converge toward classical trajectories even in
the regime where $\hbar$\ becomes negligibly small relative to the
classical action (or analogue quantities having dimensions of an
action). It is therefore interesting to examine how the de
Broglie-Bohm theory accounts for quantum phenomena in which the
manifestations of the classical dynamics is apparent, such as the
phenomena portrayed in Fig.\ 1 for the hydrogen atom in a magnetic
field.

\begin{figure}[tb]
\begin{center}
\includegraphics[height=2.6in,width=3.6in]{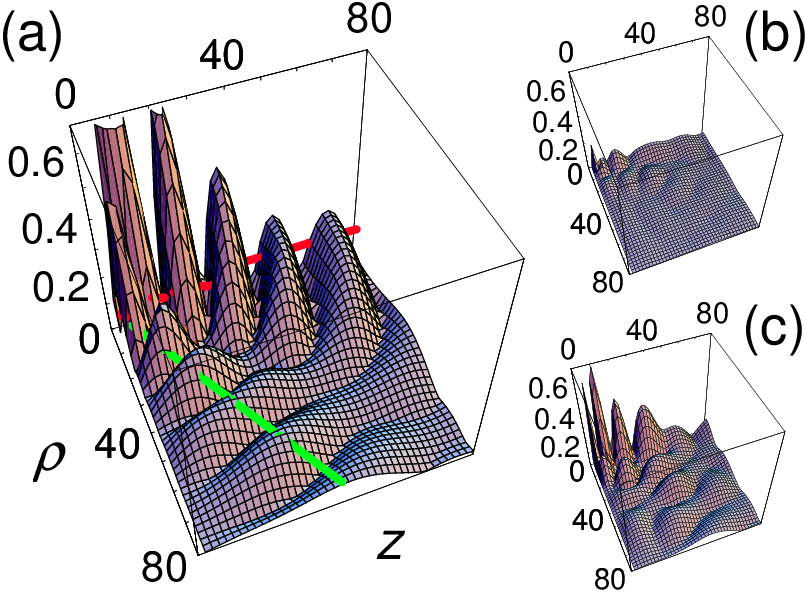}
\caption[]{The wavefunction $|\psi(\varrho,z,t)|$ in the region near
the nucleus (restricted to the fundamental quadrant, coordinates in
atomic units). (a) The initial wavefunction (a radial Gaussian
centered at $r_0=10$ and an angular distribution in the form of  a
double bump with maxima at $\theta_B$ and $\theta_C$) is shown
shortly after $t=0$, when the wavepacket starts to propagate.
$|\psi|$ is given in units of the highest value of this quantity
appearing in the graph. We have also represented 2 trajectories
having the same initial position $\mathbf{r}(t=0)=(10,\theta_C)$:
the \emph{classical} trajectory (green line leaving the nucleus
region with a straight angle $\theta_C$) and the \emph{BB quantum}
trajectory (red line, which drifted by the main probability flow
immediately turns left toward the $z$ axis). (b) The wavefunction
near the nucleus at $t=8.9$ ps, when most of the wavepacket has left
this region. The vertical scale is the same as in (a). (c) Same as
(b) at $t=16.4$ ps when part of the wavepacket returns to the
nucleus region, giving rise to the first peak in the autocorrelation
function (see Fig. 3).} \label{figinit}
\end{center}
\end{figure}

We will examine some quantum trajectories obtained from the time
evolution of an initially chosen wavefunction. In this work we will
only be interested in the specific case of recurrences; a full
account detailing the global properties of the quantum trajectories
for the H atom in a magnetic field problem, in particular as a
function of the classical dynamical regime, will be given elsewhere
\cite{matz-prep}. To compute quantum trajectories, we numerically
integrate Eq. (\ref{e10}) in the axial $(\varrho,z)$ plane.\ This is
computationally very demanding.\ First, an initial wavefuntion
localized in configuration space results from an expansion of
several hundred energy eigenstates $\psi_{E_{k}}(\mathbf{r})$.\
Second, each of these eigenstates contains several hundred thousand
components on the numerical basis, obtained from the diagonalization
of very large (but sparse) Hamiltonian matrices \cite{friedrich
wintgen,delande94}.\ Third the integration of Eq. (\ref{e10}), where
$\mathbf{\triangledown}\sigma$ is obtained from the logarithmic
derivatives of the wavefunction, calls for very small steps when
approaching the nodes of the wavefunction. The complexity of the
nodal pattern depends on different parameters \cite{matz-prep}, but
the number of nodes increases at least quadratically with the
excitation energy. The compromise made was to work with moderately
excited states and an initial state that is only approximately
localized : in the expansion
\begin{equation}
\psi(\mathbf{r},t=0)=\sum_{k}\alpha_{k}(\mathbf{r}_{0})\psi_{E_{k}}%
(\mathbf{r})\simeq ce^{-(\mathbf{r}-\mathbf{r}_{0})^{2}/2\Delta r}\label{22}%
\end{equation}
the sum over $k$ is restricted so that the Gaussian on the right
handside is only approached. Note that besides the axial symmetry,
the Hamiltonian (\ref{5}) possesses a definite symmetry by
reflection on both the $z$ and $\varrho$ axis: it is therefore
sufficient to consider the upper right 'fundamental' quadrant
($\varrho$ and $z$ both positive; in polar coordinates $\theta$
refers to the angle with the $z$ axis and varies from 0\ to
$\pi/2$).

\begin{figure}[tb]
\begin{center}
\includegraphics[height=2.75in,width=1.8in]{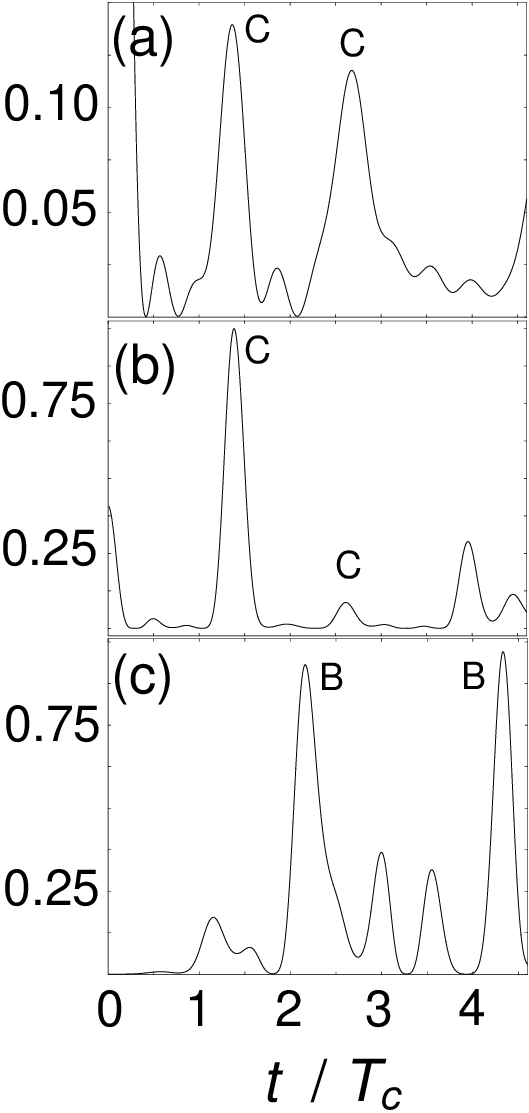}
\caption[]{Time variation of  the autocorrelation function
$|C(t)|^{2}$ (a), the local probability density
$|\psi(r^{\prime},\theta _{C},t)|^{4}$ (b)
$\left|\psi(r^{\prime},\theta _{B},t)\right|^4 $ (c), for the
initial wavefunction described in the text ($r^{\prime}=60$ au). The
time is given in units of the cyclotron period $T_c=11.8$ ps and the
vertical scale is normalized to the highest value appearing in the
graph (for the autocorrrelation function (a), this means that the
recurrences account for about $\sqrt{0.14} \simeq 37$\% of the
initial probability). The letter on a peak refers to the orbit whose
period matches the time of the peak maximum (the shape of the
relevant orbit is given in Fig. 1).}

\end{center}
\end{figure}

\section{Results}

We take an initial state of the form (\ref{22}) localized near the
nucleus with a radial Gaussian peaking at $r_{0}=10$ au and an
angular distribution in the form of double bump with maxima at
$\theta_{B}=0$ and $\theta_{C}=1.1$ (see Fig. 2). The initial angles
are chosen (and therefore labeled) so that they correspond to the
angles of the outgoing classical orbits B and C drawn in Fig. 1. The
value of the magnetic field is fixed at $B=3$ T and the mean energy
of the wavepacket corresponds to $n=55$ (yielding a mean scaled
energy of $\epsilon=-0.3$). About 60 states are included in the sum
(\ref{22}), with $n$ in the range 53 to 58 (therefore $\epsilon$ is
not strictly \emph{fixed} like in the case portrayed in Fig.\ 1). As
$t$
increases the wavepacket propagates according to%

\begin{equation}
\psi(\mathbf{r},t)=\sum_{k}e^{-iE_{k}t}\psi_{E_{k}}(\mathbf{r})\left\langle
\psi_{E_{k}}\right|  \left.  \psi(t=0)\right\rangle
\end{equation}
and reaches regions several thousand of atomic units away from the
nucleus. However part of the wavepacket returns to the nucleus: this
is readily visible
on the autocorrelation function%
\begin{equation}
C(t)=\left\langle \psi(t=0)\right|  \left.  \psi(t)\right\rangle =\sum
_{k}\left|  \alpha_{k}\right|  ^{2}e^{-iE_{k}t}%
\end{equation}
or by simply monitoring the probability density $\left|  \psi(r_{0}%
,\theta,t)\right|  ^{2}.$ Such quantities are plotted on Fig.\ 3.
(a) shows $\left|  C(t)\right|  $, (b) gives $\left|
\psi(r^{\prime},\theta _{C},t)\right|  $ and (c) $\left|
\psi(r^{\prime},\theta_{B},t)\right|  $ where $r^{\prime}\simeq 60$
au is chosen sufficiently far from the nucleus so that the
wavepackets converging toward the nucleus from different directions
are sufficiently well spatially separated.

The important feature seen in Fig.\ 3 concerns the presence of
isolated peaks. These peaks correspond to recurrences of the
wavepacket. These recurrences appear at times correlating with the
periods of the closed classical orbits B and C shown in Fig. 1 (the
periods are found by numerically integrating the classical equations
of motion along the chosen orbit, from which it is found that C has
a classical period of 16.3 ps, and B a period of 25.4 ps)
\footnote{The classical orbits closed at the nucleus are obtained by
a numerical integration of the classical equations of motion coupled
to a root-finding procedure by varying $\theta$ while $r_{0}$ and
$p_{\theta}=0$ are kept fixed \cite{delos88}.}.
\begin{figure}[tb]
\begin{center}
\includegraphics[height=2.75in,width=4in]{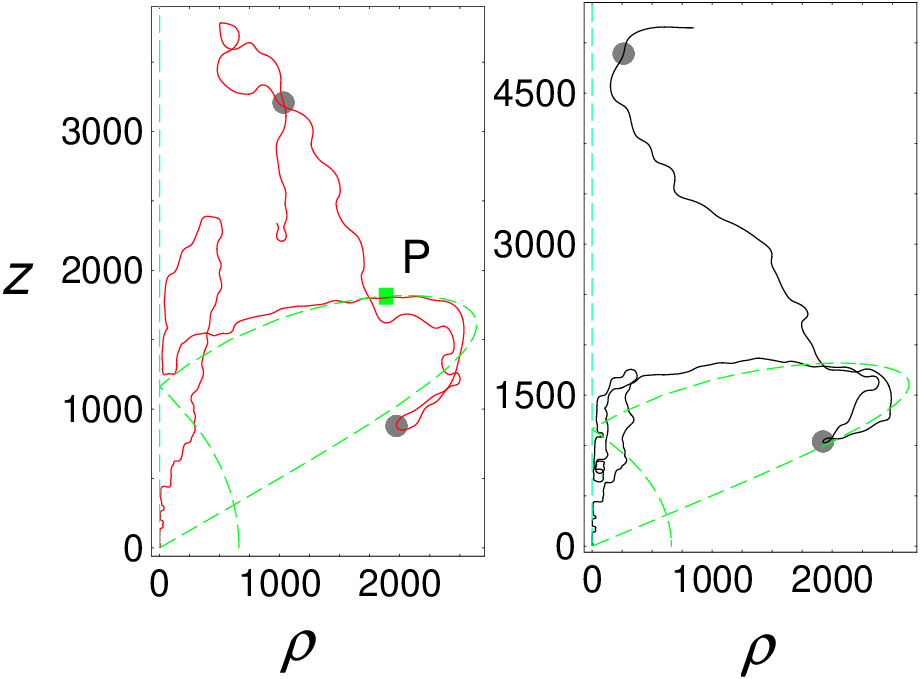}
\caption[]{Quantum trajectories (solid lines) in the $(\varrho,z)$
plane, as determined by BB theory when the initial wavefunction is
localized near the nucleus ($r_0=10$, two bumps at $\theta_C$ and
$\theta_B$). Left panel: initial position of the trajectory at
$\mathbf{r}(t=0)=(10,\theta_C)$, right panel initial position at
$\mathbf{r}(t=0)=(10,\theta_B)$. The dotted lines show the classical
trajectories B and C of Fig. 1, leaving the nucleus region with
respective angles $\theta_B$ and $\theta_C$, in the fundamental
quadrant (B ([blue dashed] turns back at $z=6000$ a.u., outside the
range of the plots; in the fundamental quadrant, C [green-dashed]
bounces on the $z$ axis changing branches and is reflected backward
by the $\varrho$ axis [compare with the overall shape of the orbit
shown in Fig. 1]). The grey dots correspond to the position of the
electron along the BB trajectory at times corresponding to the high
peaks in the autocorrelation function (Fig. 3a), i.e. the period and
twice the period of the classical trajectory C. The square on the
left panel locates the point P (see Fig. 6) }

\end{center}
\end{figure}
The quantum trajectories with initial positions at the maxima of the
bumps $\mathbf{r}(t=0)=(10,\theta_C)$ and
$\mathbf{r}(t=0)=(10,\theta_B)$ (BB$_C$ and  BB$_B$) are shown in
Fig. 4 (see also Fig.\ 2 for a closeup of BB$_C$ near the nucleus).
The shape of both trajectories is at first similar, although BB$_B$
spends more time near the $z$ axis. It is readily apparent that
neither BB$_C$ nor BB$_B$ come back to the region near the nucleus
at times compatible to account for the recurrence peaks seen in Fig.
3 (the positions of the Bohmian particle at the time of the first
two peaks seen in the autocorrelation function is shown in Fig 4).

Fig. 5 shows other typical examples of quantum trajectories. The
left panel shows quantum trajectories for the same initial
distribution as in Fig. 4 but different initial positions. The right
panel shows the quantum trajectories with the same initial
conditions as in Fig. 4 but for a slightly different initial
wavefunction: in the sum defining the initial wavefunction in Eq.
(\ref{22}), we have left out the first and the last terms. If we
would have plotted the new initial wavefunction as in Fig. 2, the
difference would not be visible to the eye; neither would the time
dependent functions shown in Fig. 3(a) and (b)  be different on the
scale of the plots (Fig. 3(c) would be barely different). However
the BB trajectories plotted in Fig. 5 (right) are markedly different
from their counterpart of Fig. 4.

\section{Discussion and conclusion}

We have seen for a specific initial wavefunction the presence of
quantum recurrences appearing at times matching the period of
classical orbits closed at the nucleus. As mentioned above, this
behaviour has been experimentally observed in the recurrence spectra
of hydrogen and other Rydberg atoms in a magnetic field
\cite{friedrich wintgen,mainprl,delande94}. The dynamical
interpretation of such experiments and of the results shown in Fig.
3 relies on semiclassical arguments: most of the initial
wavefunction was chosen to sit near the nucleus in regions
overlapping with the classical trajectories B and C. Since according
to Eq. (\ref{18}) the wavefunction is carried by the classical
trajectories, a recurrence seen at time $T$ corresponds to the part
of the wavepacket that returns to the nucleus along the classical
trajectory with period $T$. According to these semiclassical
arguments, the wave simultaneously travels along the available paths
-- this is an instance of a sum over paths, not an application of
the Erhenfest theorem. If this sum over paths picture makes sense,
then by monitoring the probability amplitude at some space point
along a chosen orbit, one should detect the passage of the
wavepacket at times compatible with classical motion. This is indeed
the case, as shown in the example given in Fig. 6 for the point P
chosen on the closed orbit C. Classically, the travel time from the
initial position $\mathbf{r}(t=0)=(10,\theta_C)$ to P is about
$0.43$ $T_c$; the classical particle continues to travel along C,
reflects on the $\varrho$ axis and travels backward, reaching P
about $0.48$ $T_c$ later. The first 2 recurrence peaks seen in Fig.
6 agree with these classical times. The two last peaks on the right
appear at times reflecting the shift of the first 2 peaks by one
classical period of the orbit C, about $1.4$ $T_c$.

\begin{figure}[tb]
\begin{center}
\includegraphics[height=2.75in,width=4in]{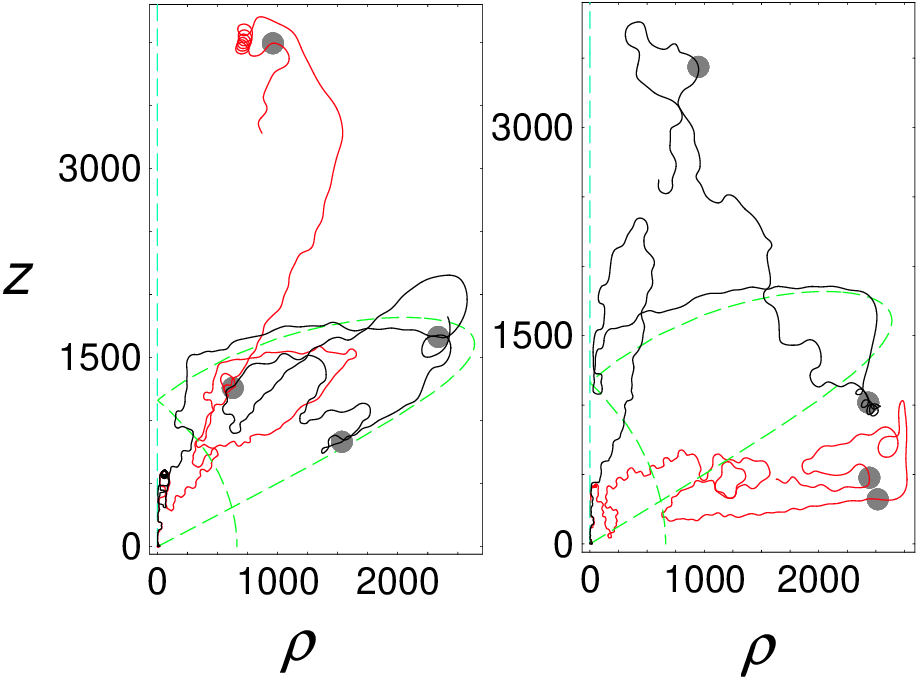}
\caption[]{Quantum trajectories. Left panel: same as Fig. 4 but with
the following initial conditions: $\mathbf{r}(t=0)=(5,\theta_C)$
(red line) and $\mathbf{r}(t=0)=(9,0.25)$ (black line). Right panel:
same as Fig. 4 [$\mathbf{r}(t=0)=(10,\theta_C)$ (red) and
$\mathbf{r}(t=0)=(10,\theta_B)$ (black)] but when the initial
wavefunction is slightly different (see text); although the
wavefunction dynamics is globally  almost identical in both cases,
the BB trajectories are different.}

\end{center}
\end{figure}

As mentioned in Sec. 3, the de Broglie-Bohm trajectories are
directly related to the density current. As can partially be
inferred from Fig. 2, the density current is at first higher near
the $z$ axis: this is why BB$_C$ quickly turns left and escapes
along this axis. Now we have remarked that only a fraction of the
initial wavefunction actually returns to the nucleus to produce the
observed recurrences. It is therefore quite improbable that a
particle with initial position at a probability maximum leaving the
nucleus region by following the main current will return to the
nucleus to account for the observed recurrences. From this
perspective the fact that the BB trajectories initially near the
maximum of the probability distribution are very far from the
nucleus when the recurrences are produced, as testified by the dots
in Fig. 4, is not surprising. This appears to be a generic property
of BB trajectories for systems in the semiclassical regime, as
discussed in Sec. 3. Of course, needless to mention that this point
has nothing to do regarding the capacity of BB theory to account for
these recurrences. This simply entails that the partial revival of
the wavefunction seen e.g. in Fig 2(c), that translates as a peak in
the autocorrelation function, should be attributed to a Bohmian
particle that occupied at $t=0$  a point in configuration space away
from the probability maximum of the distribution. The precise
initial position of such a particle will depend on the system
specifics (current density and initial state). In the present case,
although the highest values of the initial probability distribution
are found by far near the nucleus, the probability outside this zone
is not zero (although it is several orders of magnitude below the
value near the nucleus). Hence the initial positions of the Bohmian
particle accounting for the recurrences can be found inside or
outside the nucleus region (for instance the BB trajectory arriving
exactly at $(r_0,\theta_C)$ at the period of trajectory C was at
$t=0$  near $\varrho=700$, $z=128$ au, where the initial probability
is 4 orders of magnitude less than near the nucleus). We therefore
see that on the statistical level the recurrences can be explained
by taking into account the ensemble of different initial positions,
scattered throughout all of configuration space, that lead the
particle to the nucleus region at times corresponding to the
observed recurrences. As another illustration take the recurrences
at P seen in Fig. 6: the BB trajectory BB$_C$ (having its initial
position on the bump corresponding to the initial position of the
orbit C) also goes through P (Fig. 3), reaching P at $t=0.97$ $T_c$.
Therefore this quantum trajectory can account for the second peak
seen in Fig. 6, but not for the first nor the last two peaks. BB
trajectories going through P to account for these other peaks do
exist (they all have different initial conditions). However although
according to the propagator (\ref{1}) these classical-time
recurrences are due to the propagation of a part of the wavefunction
on a classical periodic orbit, there is no such dynamical
explanation in terms of the motion of a Bohmian particle on a given
trajectory.

It is also worth noticing that is not unusual for the BB
trajectories to follow some segments of classical trajectories. For
example in Fig. 4 BB$_C$ and BB$_B$ follow at short times the B
orbit parallel to the $z$ axis, and then tend to organize around C
for some time (both BB orbits go from the $z$ axis, through P, and
turn with C). Indeed when a classical trajectory travels along the
current density gradient, a nearby BB trajectory will present  the
same motion. But as a general rule classical trajectories cross and
the net current ruling the BB motion arises from the resulting
interference (the simplest example being the one-dimensional
infinite well, where the classical to and from motion results in an
interference leading to a static net current and hence no BB motion,
e.g. Sec. 6.5 of \cite{holland93}).

The detailed motion of the Bohmian particle also depends on the
dynamics of the nodes, which may either trap the particle for a long
time or violently separate two nearby BB trajectories (examples will
be given elsewhere \cite{matz-prep}). The result is that BB
trajectories are in general considerably more complex than the
classical ones, as recently put  in evidence in the case of stadium
billiards \cite{wisniacki03}. This is why as  seen in Fig. 5  a
slight change in the initial wavefunction which barely affects the
subsequent time-evolution of observable (and statistical) quantum
quantities may give rise to very different BB trajectories. Indeed
the quantum potential is very sensitive to locally fine details of
the evolving wavefunction. Therefore two slightly different
wavefunctions may give rise, in terms of the Bohmian particle, to
different dynamics (compare the red curves in the left panel of Fig.
4 and the right panel of Fig. 5, which have the same initial
position).  On the other hand, the semiclassical arguments are the
same for both initial wavefunctions whose large-scale structures
depend on the underlying classical dynamics identical in both cases,
the relative weight of the interfering trajectories being different.

\begin{figure}[tb]
\begin{center}
\includegraphics[height=1.5in,width=1.9in]{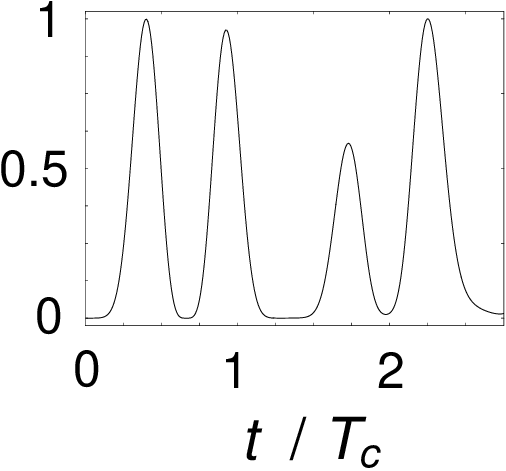}
\caption[]{$\left|\psi(\varrho_{P},z _{P},t)\right|^4$ is given as a
function of the cyclotron period $T_c$; P is a point chosen on the
closed classical trajectory C (the position of P is shown in Fig.
4).}

\end{center}
\end{figure}

To summarize, we have investigated wavepacket dynamics of the
hydrogen atom in a magnetic field when the wavepacket is initially
localized near the nucleus. This quantum system displays the
fingerprints of classical trajectories in the form of recurrences
appearing at times matching the periods of the closed orbits of the
classical system. These features are well understood within the
semiclassical approximation to the path integral propagator,
grounded on the properties of the classical trajectories. Quantum
trajectories computed according to the de Broglie Bohm theory do not
display such periodicities: individual trajectories are highly
nonclassical and cannot explain the observed recurrences. Their
dynamics is governed by the local current density which for highly
excited systems is extremely complex. The recurrences are
nevertheless accounted for statistically by the arrival at the
recurrence times of particles whose initial positions were
preferentially away from the maximum of the initial distribution.
This statistical explanation may seem to lack a dynamical
determination concerning the particle: the manifestation of the
classical motion apparent in the large scale structures of the wave
propagation, inducing the partial periodicity of the quantum system,
has no counterpart in the motion of the Bohmian particle, guided by
the local details of de Broglie's 'pilot-wave' (along the
probability current density). Possible consequences regarding the
emergence of classical trajectories from quantum mechanics in a de
Broglie-Bohm framework will be examined in a future work.

\end{document}